\documentstyle[12pt] {article}
\newcommand{\beq}{\begin{equation}}
\newcommand{\eeq}{\end{equation}}
\newcommand{\beqa}{\begin{eqnarray}}
\newcommand{\eeqa}{\end{eqnarray}}
\newcommand{\ba}{\begin{array}}
\newcommand{\ea}{\end{array}}

\begin{document}

\begin{center}
{\large \bf Chaotic features in classical scattering processes \\
between ions and atoms}
\vskip 0.5cm
{\bf Fabio Sattin}${}^\dagger$
{\footnote{E-mail: sattin@pdigi3.igi.pd.cnr.it}}
and {\bf Luca Salasnich}${}^\ddagger$
{\footnote{E-mail: salasnich@math.unipd.it}}
\vskip 0.3cm
${}^\dagger$ Dipartimento di Ingegneria Elettrica, Universit\`a di Padova, \\
via Gradenigo 6/a, 35131 Padova, Italy  \\
${}^\ddagger$ Dipartimento di Matematica Pura ed Applicata, 
 Universit\`a di Padova\\
via Belzoni 7, 35131 Padova, Italy \\
and \\
Istituto Nazionale di Fisica Nucleare, Sezione di Padova, \\
via Marzolo 8, 35131 Padova, Italy 
\end{center}
\vskip 0.5cm
{\bf Abstract}
A numerical study has been done of collisions between protons and 
hydrogen atoms, treated as classical particles, 
at low impact velocities. The presence of chaos has been looked for
by investigating the processes with standard 
techniques of the chaotic--scattering theory. 
The evidence of a sharp transition from nearly regular scattering to fully 
developed chaos has been found at the lower velocities.
\vskip 0.5cm
PACS numbers: 05.45.+b, 34.70+e

\newpage

Since a long time the use of classical models to study collision processes 
between atomic particles has been one of the main tools of 
atomic physics, and among these methods a major role has been 
played by the Classical Trajectory Monte Carlo (CTMC) method. 
In this method, originally conceived to study collisions involving an hydrogen
atom and a fully stripped ion [1], all the particles 
involved are treated classically. Randomness is introduced at the level of 
initial conditions of the motion: electron coordinates are picked up from 
statistical distributions aiming to mimic quantum ones. 
At the end of the collision process the electron may be still bound 
to the original nucleus, possibly in a different quantum state 
(excitation of the target), it may be bound to the 
projectile (charge transfer), or may simply be ionized. The physical 
quantities of interest characterizing the collision ({\it i.e.} the cross 
sections for each process) are obtained by averaging the outcomes over 
the initial statistical distribution. 
Along the years the CTMC has been extended and refined to yield an increasing
amount of informations ({\it e.g.} differential 
cross sections) as well as to include more complex systems such as 
multi-electron atoms. For surveys about the subject one can see [2] and 
references therein. A main limitation of CTMC is that its validity is 
theoretically justified only for collision energies not too low 
(the natural unit of measure for velocities in this system is given by the 
electron classical orbital velocity $v_e$, which is the Bohr velocity, or 
about 2.2 $10^6$ m/s, for an H atom in its ground state). 
In defect of this criterion quantum effects are thought to deeply influence
the dynamics through tunnelling of the electron between the nuclei. Since 
the method shows great simplicity and easiness of use if compared to fully 
quantal calculations it is tempting to extend it also in this "forbidden" 
region. Recently Keller {\it et al} [3] have shown that 
quantum effects in the dynamical evolution of the electron can satisfactorily 
be simulated by a stochastic algorithm which adds small random gaussian 
fluctuations to its otherwise deterministic trajectory. 
It is shown that: i) the new algorithm improves the results of the standard 
CTMC as compared to experiment and to fully quantum mechanical  calculations 
at low energies, and ii) the final results are quite insensitive to the
details of the perturbation. 
\par
However, in spite of the practical effectiveness of the method, yet no sound 
theoretical reason exists to account for this robustness. 
Recently, Katsonis and Varvoglis [4] have suggested that the key may be found 
in the structure of the phase space of the system: if all or a great part of it
is {\it chaotic} then it is known that such a system is {\it stable} against 
perturbations, namely small modifications of the Hamiltonian do not change 
the qualitative behaviour of its dynamics. As long as quantum mechanical 
features can be seen as "perturbations" to the classical Hamiltonian, the same 
argument justifies the effectiveness of the classical method in front of the 
experimental results. The validity of this argument relies on the phase space 
to be entirely or mostly chaotic. If large regions of ordered motion exist 
then perturbations can destroy them and this will be reflected into final 
averages. In [4] it is suggested that this hypothesis may be verified by 
numerically investigating the phase space in search of chaotic regions. 
The scope of this paper is exactly to  start an analysis into this 
direction. 
\par
Since the dimension of the phase space is rather large a complete study is 
difficult to perform and we have not attempted it: instead of a systematic 
investigation we have carried out a set of calculations searching for 
presence of traces of chaos. 
For this reason we have also partially sacrificed the physical 
soundness of the model by inspecting the region of very low impact energies 
($v \ll 1$) where chaos is more likely to be found but where the classical 
approximation loses most of its validity. It has to be remarked
that even to get the results displayed below it took several hundred hours 
of CPU time on a DEC ALPHA 2100 workstation. 
\par
In the following we will use atomic units, where the velocities are measured 
in units of the Bohr velocity $v_e$.
\par
The system is initially prepared with the target nucleus, a proton, 
in the origin of coordinates with null velocity, 
and with the electron orbiting around it. 
Initial electron coordinates (position and momentum) satisfy the condition 
that its total energy must be $ - 0.5$ units 
(the binding energy of a 1s electron). The proton 
projectile must start from a large distance ($>10$ units in our runs) 
in order that its initial interaction with the target 
be considered negligible. The equations of the motion are numerically 
integrated using a fourth order Runge--Kutta--Merson method [5] until the 
two nuclei are well far apart. The final state 
of the electron---ionized, captured from the projectile, or simply excited--- 
is determined from standard procedures [1,6]. 
The problem is a standard one of chaotic scattering: we have a dynamical 
system free to evolve from a given initial state 
(which is characterized by an input variable, or set of variables, $\theta_i$), 
and look for the value of the final state $\theta_f$ as a function of 
$\theta_i$. 
Usually $\theta_f$ is a smooth function of $\theta_i$ apart possibly a number 
of singular points: 
in any arbitrarily small neighborhood of the singularities the 
output variable varies wildly. The chaotic regions of the phase space are 
responsible for the singularities: while always of Lebesgue measure zero, 
the set of singularities has a fractal dimension greater 
than zero in correspondence of a chaotic region (for a 
complete discussion about the argument, see the reviews of Eckhardt [7] 
and of Ott and T\'el [8]). 
\par
The determination of the fractal dimension is based upon the uncertainty 
exponent technique developed by Bleher {\it et al} [9,10]. 
In this work we have kept fixed the initial electron coordinates 
(${\vec r} , {\vec p}$), the initial internuclear distance and the relative 
speed, and the only free parameter left is the impact parameter $b$.
By this procedure the dimension of the phase space is reduced to one. The 
system is let to evolve and the outcome is recorded. Then the  procedure is  
repeated using a slightly different impact parameter, $b \pm \epsilon$, and 
the results of the two runs are compared. The final state is said uncertain 
--after Bleher {\it et al} [9]--if the two results are different. 
If we label each of the possible outcomes with an integer number, and plot 
them {\it vs} $b$, in correspondence of an 
uncertain trajectory we have a discontinuity of the output variable. Some 
plots of this kind are shown in Figure 1. 
\par
We have computed the fraction $ f(\epsilon) $ of uncertain 
trajectories as a function of the parameter $\epsilon$: it is known that 
\beq
f(\epsilon) \propto \epsilon^{\alpha} \;  \;\;\;\;\;\; 
\hbox{for} \;\; \epsilon \to 0 \; , 
\eeq
and $\alpha $ is related to the boundary fractal 
dimension (capacity dimension) $d$ through $\alpha = D - d$, where $D$ is the 
dimension of the parameter space (here, $D = 1$) [9,10]. 
If the scattering is regular $d = 0$ and $\alpha = 1$, 
while in presence of fully developed chaos $d = 1$ and $\alpha = 0$.

We have first made a scan over impact parameter for different velocities 
in order to identify the chaotic regions. In Figure 1 the results are plotted 
for one value of $v$ and zooming on $b$. We have arbitrarily assigned the 
value 1 if at the end of the collision the electron remains bound to the 
target nucleus, 2 if it is ionized, and 3 if it is bound to the projectile 
nucleus.It is quite clear that a kind of self--similarity appears. 
Similar plots at nearly equal velocities show all the same features which 
disappear at higher values of $v$. 
\par
Figure 2 shows $f(\epsilon)$ {\it vs} $\epsilon$ for some 
different velocities. 
At the lower velocities two different regions are quite clearly visible: 
one, for $\epsilon \to 0$, where there is a clear power--law behaviour, 
and another, at greater $\epsilon$. 
The correction at great $\epsilon$ is due to a logarithmic term [11],
\beq
f(\epsilon ) \propto \epsilon^{\alpha} (\ln({1 \over \epsilon}))^{\beta}  
\eeq
which does not affect the determination of the capacity dimension. In [11] a 
similar behaviour has been associated with a case of {\it nonhyperbolic } 
chaotic scattering, in which the fractal dimension is $d = 1$. \\
In the calculation of the capacity dimension we have fitted the data with 
simple power--law curves, and skipped the range of $\epsilon$ where the 
power--law behaviour is not dominant. 
\par
In Figure 3, which represents the main result of this work, $d$ is plotted 
{\it vs} impact velocity. One observes a sharp transition between a situation 
of nearly absence of chaos and one of fully developed chaos at a velocity 
$v_c$ of about 0.06 units. Within statistical errors we cannot state whether 
there is or not chaos beyond $v_c$: further studies are presently being 
carried on. 
This situation resembles that illustrated in [9,11-13] of abrupt transitions 
from a scattering regime to another. 
From Figure 2 we see that the transition to non chaotic 
scattering is associated with an increasing of the crossover region, 
where the logarithmic term in equation (2) is not negligible. 
On the contrary, at higher velocities, above the transition, no logarithmic 
behaviour is visible. 
The critical value of the transition corresponds to an impact energy of 
about $E=100$ eV, 
but we are still not able to associate a clear meaning to this energy. 
The value of $v_c$ is much below that studied in [3] ($ v_c > 0.6$) so the 
present results cannot support the hypoteses of [4]. 
However presence of chaos at higher velocities cannot absolutely be 
excluded on the basis of our investigations. 
\par
The phase space associated to electronic coordinates obviously plays a 
crucial role in determining the dynamics of the scattering. We have done
several simulations in this sense, varying the initial 
coordinates (${\vec r}, {\vec p}$); our computations have not allowed us to 
see any difference: the functional dependence is the same even if absolute 
values of $f$ are different.
\par 
From our numerical calculations we can conclude that there is chaos in 
classical processes of charge exchange: it appears in the form of a sudden 
transition from fully developed chaos to regular motion when increasing the 
velocity.  
\par
We remark that even if the original motivation for this work was about 
the connections between chaos and the CTMC method, our final results cannot 
have a direct relevance to CTMC due to the consideration of low impact
velocities.
\par
Finally, we point out that up to now only the classical theory has been dealt 
with but some of the 
results previously shown may be useful in the quantum theory of 
chaos: actually the $H$--proton scattering at these low energies is perfectly 
amenable to quantum calculations so we have an example of quantum 
system whose classical counterpart exhibits chaos, 
the study of these systems being an active area of research. 

\vskip 0.5cm

F.S. has been financially supported during this work by a grant of the 
Italian MURST. Computing facilities have been made available by the Istituto 
Gas Ionizzati del CNR of Padova. 

\newpage

\section*{References}

\parindent=0.pt

[1] R. Abrines and I.C. Percival, Proc. Phys. Soc. {\bf 88}, 861 (1966); 
R. Abrines and I.C. Percival, {\it ibid.} p. 873

[2] P.T. Greenland, Phys. Rep. {\bf 81}, 131 (1982); B.H. Bransden and M.R.C. 
McDowell, {\it Charge Exchange and the Theory of Ion--Atom Collisions}, Oxford 
Science Publications (1992), chap. 8

[3] S. Keller, H. Ast and R.M. Dreizler, 
J. Phys. B: At. Mol. Opt. Phys. {\bf 26}, L737 (1993)

[4] K. Katsonis and H. Varvoglis, 
J. Phys. B: At. Mol. Opt. Phys. {\bf 28}, L483 (1995)

[5] Subroutine D02BAF, The NAG fortran Library, Oxford

[6] K. T\"okesi and G. Hock, 
Nucl. Instrum. Meth. Phys. Res., Sect. B {\bf 86}, 201 (1994)

[7] B. Eckhardt, Physica D {\bf 33}, 89 (1988)

[8] E. Ott, T. T\'el, CHAOS {\bf 3}, 417 (1993)

[9] S. Bleher, C. Grebogi, E. Ott, and R. Brown, Phys. Rev. A {\bf 38}, 930 
(1988); 
S. Bleher, E. Ott and C. Grebogi, Phys. Rev. Lett., {\bf 63}, 919 (1989);
S. Bleher, C. Grebogi and E. Ott, Physica D {\bf 46}, 87 (1990)

[10] C. Grebogi, S.W. McDonald, E. Ott, and J.A. Yorke, Phys. Lett. A {\bf 99}, 
415 (1983); 
S.W. McDonald, E. Ott, and J.A. Yorke, Physica D {\bf 17}, 125 (1985)

[11] Y.-T. Lau, J.M. Finn, and E. Ott, Phys. Rev. Lett. {\bf 66}, 978 (1991)

[12] Y.-C. Lai, C. Grebogi, R. Bl\"umel, and I. Kan, 
Phys. Rev. Lett. {\bf 71}, 2212 (1993)

[13] M. Ding, C. Grebogi, E. Ott, J.A. Yorke, Phys. Rev. A {\bf 42}, 
 7025 (1990)

\newpage
\section*{Figure Captions}
\vskip 0.3cm

{\bf Figure 1}: scattering outcome {\it vs } impact parameter $b$. Results 
are labelled according to: excitation, 1; ionization, 2; charge transfer, 3. 
From the top to the bottom successive blowups are shown. 
Respectively 1000, 600, and 700 points have been plotted.
\\
{\bf Figure 2}: plot of $f(\epsilon)$ {\it vs} $\epsilon$. Squares: 
$v = 0.0282$; stars: $v = 0.0447$; triangles: $v = 0.2$. 
\\
{\bf Figure 3}: capacity dimension $d$ {\it vs} impact velocity $v$. 
Error bars are the errors on the slope of the straight 
lines $\ln{f}$ {\it vs} $\ln{\epsilon}$ as calculated by linear fits.

\end{document}